\documentclass[12pt]{iopart}

\usepackage{graphicx}
\usepackage{graphics}
\usepackage{appendix}
\usepackage{multirow}
\usepackage{hhline}
\usepackage[normalem]{ulem}
\usepackage[usenames, dvipsnames]{color}
\usepackage{color}
\usepackage{amssymb}
\usepackage{cite}

\usepackage{hyperref}
\hypersetup{
    colorlinks=true,
    linkcolor=blue,
    filecolor=blue,      
    urlcolor=blue,
    citecolor=blue
}

\newcommand{\red}[1]{#1}
\renewcommand{\sout}[1]{\unskip}
\newcommand{\redd}[1]{#1}
\newcommand{\soutt}[1]{\unskip}

\newcommand{\myWidth}{0.65\textwidth}
\newcommand{\rd}{\mathrm{d}}
\newcommand{\eqref}[1]{(\ref{#1})}
\newcommand{\D}{\emph{D}}

\newcommand{\mt}{\tau_{50}}
\newcommand{\tmin}{\Delta t_{\rm min}}
\newcommand{\tapx}{\widetilde{\tau}}
\newcommand{\Dt}{\Delta t}
\renewcommand{\P}{\mathrm{P}}
\renewcommand{\D}{\mathrm{D}}
\newcommand{\Pd}{\P_\D}
\newcommand{\Pdo}{\P_{\D 0}}
\newcommand{\td}{t_\D}
\newcommand{\tc}{\Delta t_{\rm c}}
\newcommand{\tfit}{\Delta t_{\rm fit}}
\newcommand{\tW}{\Delta t_{\rm W}}
\newcommand{\Ip}{I_{\rm p}}
\newcommand{\Bo}{B_0}
\newcommand{\Ro}{R_0}
\newcommand{\ts}{\Delta t_{\rm s}}
\newcommand{\paren}[1]{\left(#1\right)}
\renewcommand{\brack}[1]{\left[#1\right]}

\newcommand{\dPodt}{\frac{\rd \Pdo}{\rd t}}
\newcommand{\teff}{\tau_{\rm eff}}
\newcommand{\tE}{\tau_{\rm E}}

\begin{document}



\title[Survival analysis for disruption prediction]{\redd{An application of survival analysis to disruption prediction via Random Forests}}

\author{R.A.~Tinguely\footnote[1]{Author to whom correspondence should be addressed: rating@mit.edu}, K.J.~Montes, C.~Rea, R.~Sweeney, and R.S.~Granetz}

\address{Plasma Science and Fusion Center, Massachusetts Institute of Technology, Cambridge, MA, USA 02139}


\begin{abstract}

One of the most pressing challenges facing the fusion community is adequately mitigating or, even better, avoiding disruptions of tokamak plasmas. However, before this can be done, disruptions must first be predicted with sufficient warning time to actuate a response. The established field of survival analysis provides a convenient statistical framework for time-to-event (i.e. time-to-disruption) studies. This paper demonstrates the integration of an existing disruption prediction machine learning algorithm with the Kaplan-Meier estimator of survival probability. Specifically discussed are the \soutt{implications of} \redd{implied warning times from} binary classification of disruption databases and the interpretation of \redd{output signals from} Random Forest \redd{algorithms} \soutt{output signals} trained and tested on these databases. This survival analysis approach is applied to both smooth and noisy test data to highlight important features of the survival and hazard functions. In addition, this method is \soutt{shown to be successful in predicting the disruption of an example} \redd{applied to three} Alcator C-Mod plasma discharges \redd{and compared to a threshold-based scheme for triggering alarms. In one case, both techniques successfully predict the disruption; although, in another, neither warns of the impending disruption with enough time to mitigate. For the final discharge, the survival analysis approach could avoid the false alarm triggered by the threshold method. Limitations of this analysis and} opportunities for future work are also presented.


\end{abstract}



\section{Introduction}\label{sec:intro}
Plasma disruptions in tokamaks pose a serious risk to current experiments and future fusion devices. During a disruption, the total thermal and magnetic energies---upwards of tens \redd{or hundreds} of megajoules in future devices \redd{like SPARC \cite{greenwald2018} or ITER \cite{lehnen2015}}---can be dissipated in tens of milliseconds or \soutt{less} \redd{fewer}, leading to (i) high heat fluxes on plasma-facing components, (ii) large induced eddy and halo currents in the surrounding vacuum vessel, and (iii) generation of highly-relativistic ``runaway'' electrons. Ideally, plasma disruptions should be avoided altogether, thereby preventing damage and continuing operation. Although in reality, many disruptions will need to be mitigated; that is, the plasma will be terminated in such a way---e.g. through massive gas or shattered pellet injection---so as to minimize both damage and delay in operation. 

However, for both avoidance and mitigation, an impending disruption must first be \emph{predicted} with enough time to actuate an appropriate response. There are many past and ongoing efforts to develop disruption prediction algorithms, including \redd{work on tokamaks ADITYA~\cite{sengupta2000,sengupta2001}, Alcator C-Mod~\cite{rea2018ppcf,montes2019}, ASDEX-U~\cite{pautasso2002,windsor2005,aledda2015}, DIII-D~\cite{wroblewski1997,rea2018fst,rea2018ppcf,rea2019nf,montes2019}, EAST~\cite{montes2019}, JET~\cite{cannas2004,windsor2005,cannas2007fed,murari2008,murari2009,ratta2010,deVries2011,vega2013,cannas2014,ratta2014,murari2018,pau2018}, JT60-U~\cite{yoshino2003,yoshino2005}, J-TEXT~\cite{wang2016,zheng2018}, and NSTX~\cite{gerhardt2013}, among others.} At present, most predictors estimate the plasma \emph{state} at each moment in time, usually identified as either \emph{non-disruptive} or \emph{disruptive}. While this is an important and necessary step toward accurate disruption prediction, it is not sufficient to know, even with 100\% accuracy, that a disruption is currently occurring. Instead, the ultimate predictive capability is to estimate the future probability of a disruption. \soutt{In other words,} The goal is to answer the question, what is the probability that the plasma will survive a time $\Dt$ into the future? From this prediction, combined with knowledge of plasma actuators and mitigation systems, an attempt can be made to navigate the plasma state away from disruptive territory, or mitigation \soutt{is} \redd{would be} employed.

A framework for predicting events and estimating survival times already exists in the area of statistics called \emph{survival analysis}. Decades of research in such fields as medicine, engineering, and sociology have utilized and refined these tools and techniques. The fusion community has an opportunity to leverage this established knowledge base to confront \soutt{one of the greatest challenges facing future fusion power plants:} \redd{the challenge of} disruptions. The present work is \emph{not} the first application of survival analysis for event prediction in the realm of fusion. Only recently, ``direct hazard modeling'' was used to study and predict the onset of neoclassical tearing modes \cite{olofsson2018PPCF,olofsson2018FED} which can often lead to disruptions of tokamak plasmas. The authors of \cite{olofsson2018PPCF} actually note the \soutt{usefulness} \redd{potential} of the survival analysis approach for disruption prediction. 

The aim of this work is to demonstrate the integration of \emph{existing} disruption prediction techniques\soutt{---specifically a Random Forest (RF) \cite{breiman2001} algorithm, with real-time prediction capabilities, trained and tested on binary-classified data---} into the survival analysis framework. \redd{This work focuses on the Disruption Predictor using Random Forests (DPRF) \cite{montes2019} algorithm with real-time prediction capabilities, trained and tested on binary-classified data from the Alcator C-Mod tokamak \cite{marmar2009}, a high magnetic field ($\Bo$~=~2-8~T), compact ($\Ro$~=~68~cm, $a$~=~22~cm) device with plasma currents and densities of order 1~MA and $10^{20}$~m$^{-3}$, respectively. A database of disruption parameters has been compiled for over four thousand C-Mod discharges, containing over $10^5$ data points in total \cite{rea2018ppcf,montes2019}.}


\redd{This integration of an existing algorithm within a new framework is important for two reasons: First, as mentioned, most disruption predictors only consider a binary-classification of the current plasma state; thus, the relationship between a class label and the time until the disruption must be carefully considered when making time-to-event predictions with survival analysis. Second, these existing disruption prediction algorithms are the current standard against which survival analysis must be compared; this comparison is performed, in part, in the present work. The authors note here (and in section~\ref{sec:future}) that there is certainly more work to be done beyond the present analysis; the ultimate goal is a framework incorporating the evolution models of many plasma and operational parameters in a time-to-disruption prediction. However, this is outside the scope of the current paper, and} optimization of these methods is left for future work.

The outline of the rest of the paper is as follows: Section~\ref{sec:disruption} discusses binary classification of disruption databases and RF algorithms trained on those data sets. In section~\ref{sec:survival}, survival analysis is introduced and explored as an approach to disruption prediction. This methodology is then applied to both test and real experimental data in \redd{sections~\ref{sec:test} and \ref{sec:cmod}, respectively.} \redd{In section~\ref{sec:summary}, a summary of results is given. Finally, limitations of this analysis and opportunities for future work are presented in section~\ref{sec:future}.}

\section{Disruption database classification and prediction}\label{sec:disruption}

This section \sout{describes} \red{discusses} the binary classification scheme commonly used to classify data in disruption databases. \soutt{It is important to note that} The framework described here, in its current form, is \emph{only} applicable to predictors using binary classification. Future work could expand these methods to multi-class data. \soutt{Additionally,} We focus on the supervised Random Forest (RF) \cite{breiman2001} machine learning approach to disruption prediction. Most importantly, we discuss the implications of binary classification and interpretations of RF algorithm output signals. \redd{Regarding notation in this and following sections, we use $t$ for time, $\Delta t$ for time intervals or windows, and $\tau$ for times of interest in survival analysis.}

\subsection{Implication of binary classification}\label{sec:binary}

Many efforts to develop disruption prediction algorithms follow a similar methodology, summarized here: First, a set of plasma parameters---e.g. Greenwald fraction and internal inductance---and operational parameters---e.g. differences between programmed and measured values of plasma current or vertical position---are chosen based on their relevance to disruption physics as well as \redd{their} real-time measurement capability. Second, databases of these parameters are assembled for many machines and many times throughout plasma discharges, some terminated by disruptions\red{; for examples, see} \cite{wroblewski1997,pautasso2002,cannas2004,ratta2010,gerhardt2013,rea2018fst}.

\begin{figure}[h!]
    \centering
    \includegraphics[width=0.7\textwidth]{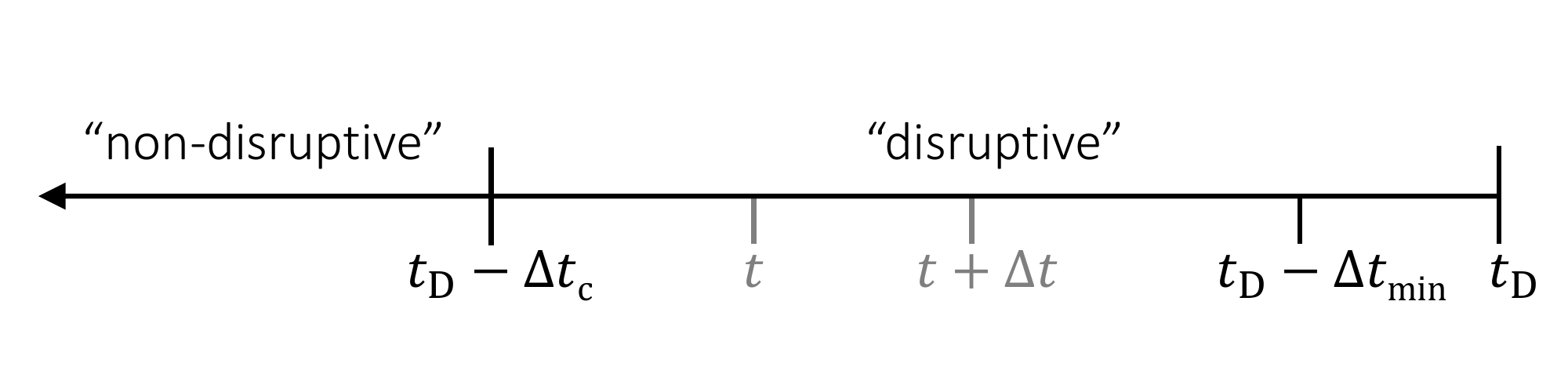}
    \caption{\redd{A representative timeline showing binary classification of a disruptive discharge with disruption time $\td$. For a class time $\tc$, data before $t < \td - \tc$ are considered ``non-disruptive,'' while data within $\td - \tc \leq t \leq \td$ are considered ``disruptive.'' Oftentimes, there is a minimum required time $\tmin$ to avoid or mitigate a disruption; therefore, an alarm must be triggered before $t < \td - \tmin$.}}
    \label{fig:timeline}
\end{figure}

\soutt{This data is} \redd{These data are} then typically split into two classes: disruptive (D) and non-disruptive (ND). All times from non-disrupting plasma discharges are considered ND, but only a subset of times from plasma discharges ending in disruptions fall in class~D. Here, it is assumed that times \emph{far from the disruption} reside in a ND region of parameter space; thus, only times \emph{close to the disruption} are classified as D. The time \redd{$\tc$} dividing the two classes \soutt{$\tc$} can be physically-motivated or optimized for best performance of the disruption prediction algorithm. In addition, due to the finite response \red{times} of the plasma control system (PCS), actuators, and \sout{the} mitigation system, data within $\td - t < \tmin$ of the disruption time $\td$ are often excluded from analyses\redd{, where $\tmin$ is \sout{specifically} the minimum time required to avoid or mitigate a disruption.} See figure~\ref{fig:timeline} for \soutt{example} \redd{an illustration of the timeline of a disruptive discharge.}

Note that in much of the previous work on this topic, \redd{e.g. \cite{cannas2007fed,gerhardt2013,ratta2014,rea2018ppcf,montes2019}}, the time evolution of parameters is not taken into account in the algorithm development; that is, each time-slice of each class is treated as independent of all other times. While time-independent studies have set up a good framework for disruption prediction, a temporal analysis is absolutely necessary for future prediction algorithms. Ultimately, \redd{it would be extremely beneficial for} the PCS \soutt{needs} to know not only the current plasma class (here D or ND), but also the probability of a disruption in the next time $\Dt$, as discussed below\redd{, in order to avoid or mitigate the disruption}.

Consider a database \redd{populated with plasma and operational parameters from many plasma discharges,} which has been split into classes ND and D by a threshold time $\tc$ before the disruption time $\td$, excluding data $\td - t < \tmin$. If a data point is randomly selected from class~D, the only information known regarding the time \emph{until} the impending disruption is that $\tmin \leq \td - t \leq \tc$, where \redd{$\td - t$ is the \emph{unknown} time interval between the time $t$ at which the data point was collected and the disruption time.}\footnote{\redd{Of course, both $t$ and $\td$ \emph{can} be determined from the database, but time dependence is \emph{not} considered in the development of these RF algorithms.}} \soutt{$t$ is the \emph{unknown} time of the given data point.} \redd{In other words, we know that each data point in class~D comes from a disruptive discharge; that is how we assembled the database, classifying data based on $\tc$. However, because our algorithm neglects the time dependence of data \emph{within} class~D, it is ignorant of the time at which each data point occurred, only knowing that $\td - \tc \leq t \leq \td - \tmin$ (refer to figure~\ref{fig:timeline}).} The probability that \soutt{this} \redd{the} random variable $t$ lies within a time interval $\Dt > 0$ \emph{before} the disruption time is given by 
    \begin{samepage}
        \begin{eqnarray}
            & 0, & \Dt < \tmin \nonumber \\
            \P(\td - t < \Dt | \D) = \Bigg\{ \; & \frac{\Dt - \tmin}{\tc - \tmin}, \; & \tmin \leq \Dt \leq \tc \label{eq:P1} \\
            & 1, & \Dt > \tc \nonumber
        \end{eqnarray}
    \end{samepage}
    
\noindent \soutt{Note that this is just another way of writing the probability that a disruption will occur in the interval $\Dt$ following any given data point in class~D, i.e. $\P(t + \Dt > \td | \D)$.}Here, the probability is conditional on the data point being in class~D, and a uniform probability density is chosen\redd{, as a conservative assumption,} due to our ignorance of the actual time until the disruption. Other probability density functions could be explored\redd{, as discussed in section~\ref{sec:future}}, but since no one \redd{option} seems particularly advantageous or physically-motivated, uniform is adopted for simplicity. \sout{Also,} \red{It} is important to note that this formulation is not ad hoc; rather, it is \soutt{purely} a consequence of the binary classification scheme. \redd{Furthermore, equation~\eqref{eq:P1} is just another way of writing the probability that a disruption occurs in the interval $\Dt$ following any given data point in class~D, i.e. $\P(t + \Dt > \td | \D)$. This is the probability of interest for real-time disruption prediction, where $t$ is known but $\td$ is not.}


From \eqref{eq:P1}, we see that our choice of $\tc$ and the value of $\tmin$ (typically machine-dependent) affect the interpretation of data within each class. Nonetheless, this formulation \soutt{at least} provides a physical interpretation of data within class~D beyond the label ``disruptive.'' \redd{That is, given any data point in class~D, the probability of a disruption occurring in the next time interval $\Dt$ increases linearly from 0 to 1 over $\tmin \leq \Dt \leq \tc$.} For simplicity in the upcoming analyses, $\tmin$ has been set to zero; this is an allowable assumption since \redd{$\tmin$} is oftentimes much less than the chosen or optimized $\tc$. \soutt{$\tmin \ll \tc$, typically.} 

It follows from \soutt{\eqref{eq:P1}} \redd{the product rule} that, given a random data point from either class~ND or D, the joint probability that the data point is in class~D \emph{and} the plasma will disrupt in the next time $\Dt$ is \soutt{simply} given as
\redd{
    \begin{equation}
        \P((t + \Dt > \td) \cap \D) = \P(t + \Dt > \td|\D) \times \P(\D) .
        \label{eq:P2}
    \end{equation}}
\noindent where P(D) is the probability of the data point belonging to class~D. The calculation of P(D) is discussed in the following section.

\subsection{Interpretation of Random Forest predictions}\label{sec:RF}

Given a disruption database split into two classes (ND and D, as described in the previous section) and an array of input parameters corresponding to one plasma state, most disruption predictors have been designed to evaluate the state's membership in either class or the distance to the ``boundary'' between classes. There are many approaches to create such algorithms. Perhaps the simplest is a \soutt{physics-based} scheme in which thresholds \redd{(of physical quantities in parameter space)} between classes ND and D are determined by minimizing overlap in \soutt{parameter} histograms \redd{of data from each class, as discussed in \cite{gerhardt2013,rea2018fst,rea2018ppcf,pau2018} among \red{other works}.} Machine learning methods have also been implemented, trained and tested on these data sets \redd{(see the references listed in section~\ref{sec:intro}).} Generally, algorithms are optimized to maximize correct classifications of states in class~D---i.e. to ``catch'' disruptions (true positives)---and minimize incorrect classifications of states in class~ND---i.e. to reduce false positives. 

This work focuses on the RF machine learning approach to disruption prediction, which has been applied on \soutt{many} \redd{several} tokamaks\redd{; see, for example, \cite{murari2008,murari2009,rea2018fst}.} In the RF framework, the parameter space of the disruption database is randomly and iteratively subdivided until the final nodes (i.e. ``leaves'' at the ends of ``branches'' of one decision ``tree'') contain instances of only one class from the training set. Many trees comprise the total ``forest.'' As new data is fed through all trees, each tree results in an evaluation of 0 (ND) or 1 (D). Thus, an average over the entire forest results in an effective probability that plasma state is in class~D. 

This output signal, called the \emph{disruptivity} in this work, is exactly the quantity of interest P(D) in \eqref{eq:P2}. At each time step, the disruptivity is calculated based on the current vector of plasma and operational parameters; therefore, the time-evolving disruptivity signal is denoted $\Pd(t)$. Real-time calculations of this signal have been demonstrated \redd{by the DPRF algorithm implemented} in the PCS system of the DIII-D tokamak \cite{rea2019nf}. Efforts are underway to perform full-discharge analyses of disruptivity signals, incorporating time evolution and optimizing both disruptivity thresholds and time-windows to trigger warnings \cite{montes2019}. \redd{These are discussed further in section~\ref{sec:cmod}.}

\section{Survival analysis}\label{sec:survival}


\red{The aim of \emph{survival analysis} is to estimate the time until a specified event---oftentimes failure or death---occurs. For instance, in medicine, survival analysis techniques are used to study patients subject to different medical treatments, importantly assessing their expected lifetime (i.e. time until death). In such a study, some data will be \emph{censored}\footnote{Specifically, this data would be \emph{right-censored}.}, whereby the event (death) will not be observed for all patients during the time of the study. This is similar, in some ways, to disruption studies as researchers often only consider data from part of the plasma discharge, such as during the flattop portion of the plasma current, before the discharge ends. However, unlike the individuals in the medical study presented above, plasmas can (effectively) be sustained indefinitely, as could be the case in a future fusion power plant. Thus, the event (here, a disruption) might never occur at all for some plasmas, no matter their discharge length.}

\red{There are many approaches within the field of survival analysis which could be applied to disruption prediction. Some attempt to determine the parametric dependence of future survival. This would be, of course, very useful from the perspective of plasma control and disruption avoidance, especially if event predictions are early enough for plasma actuation. However, disruption \emph{mitigation} is perhaps the more pressing issue since, as of yet, many disruptions cannot be predicted with adequate warning time for mitigation. Therefore, in this paper, we focus on ascertaining plasma survival probabilities and times. The survival function $S(t)$ computes the probability that the survival time $T$ surpasses a given time $t$,}     \begin{equation}
        S(t) = \P(T>t) .
    \end{equation}
The survival function is defined such that $S(t_0) \equiv 1$ at the starting time $t_0$ and $S(t)$ decreases monotonically to zero as $t \to \infty$. The corresponding hazard function $h(t)$ gives the instantaneous rate of failure, assuming survival until time $t$,
    \begin{equation}
        h(t) = -\frac{1}{S(t)} \frac{\rd S(t)}{\rd t}. 
        \label{eq:h}
    \end{equation}
\soutt{It is easy to see from \eqref{eq:h} that} If we know the hazard function over a time interval $t_a \leq t \leq t_b$, the survival function can be computed \redd{from \eqref{eq:h},}
    \begin{equation}
        S(t_b) = S(t_a) \, \exp \paren{ - \int_{t_a}^{t_b} h(t) \, \rd t }.
        \label{eq:int_h}
    \end{equation}
There are many different approaches to compute $S(t)$ for time-evolving data sets. One of the earliest and most straightforward methods is the Kaplan-Meier formalism \cite{kaplan1958}, which is adopted in this work. Given a discretized timebase, the probability of survival beyond time $t_n$ is given by the ``product-limit'' formula,
    \begin{equation}
        S(t_n) = \prod_{i=0}^n \P(T > t_i | T \geq t_i).
    \end{equation}
That is, the probability that the survival time $T$ exceeds time $t_n$ is the product of the probabilities of $T$ exceeding each time step $t_i$ before $t_n$ \emph{given} that $T$ actually reaches each $t_i$. If the probability of failure between consecutive times, \redd{i.e.} $\P_{i \rightarrow i+1}$, is known, the survival function can then be written as
    \begin{equation}
        S(t_n) = \prod_{i=0}^n \paren{ 1 - \P_{i \rightarrow i+1} }.
    \end{equation}
This is essentially equation~(7) in \cite{olofsson2018PPCF}. 

Applying the Kaplan-Meier formalism to disruption prediction is relatively straightforward: The event or failure is simply the disruption itself. For any time $t_i$, the probability of disruption in the incremental step between $t_i$ and $t_{i+1}$ is given by \eqref{eq:P2}, where $\Dt = t_{i+1}-t_i$ and $\P(\D) = \Pd(t_i)$ is the evaluation of the disruption prediction algorithm at time $t_i$. Therefore, if the current time is $t_i$, the survival probability beyond future time $t_n$ is calculated 
    \begin{equation}
        S(t_n|t_i) = \prod_{j=i}^{n} \brack{1 - \P(t_{j+1}>\td|\D)\, \Pd(t_j) } .
        \label{eq:S1}
    \end{equation}
This is written as a conditional probability because it implicitly assumes that the plasma has survived until time $t_i$. Because such a predictor would ultimately be coupled with the PCS, we assume that the disruptivity is calculated with a sampling time \redd{\emph{smaller} than the class time, i.e. $\ts < \tc$}. Then, \soutt{using \eqref{eq:P1},} the probability of survival $\Dt$ into the future from current time $t$ is
    \begin{equation}
        S(t + \Dt|t) = \prod_{j=0}^{n} \brack{1 - \Pd(t + j\ts) \frac{\ts}{\tc} } .
        \label{eq:S2}
    \end{equation}
\soutt{where $\Dt = n\ts$.} \redd{Here, the future time $\Dt$ is split into $n$ sampling times (i.e. $\Dt = n\ts$) with the constraint $\ts \leq \tc$. The disruptivity $\Pd$ is evaluated at each time step $j\ts$ from the current time $t$, and the final term $\ts/\tc$ follows from the assumption of uniform probability density in \eqref{eq:P1}.} In reality, \soutt{because} the future disruptivity signal is unknown, \redd{so} a simple linear extrapolation \soutt{could be used} \redd{is used in this work,}  
    \begin{equation}
        \Pd(t + j\ts ) \approx \Pd(t) + \frac{\rd \Pd}{\rd t} \times j \ts ,
        \label{eq:taylor}
    \end{equation}
\redd{where $\Pd(t)$ and $\rd \Pd/\rd t$ are obtained from the DPRF output.} For the purposes of this paper, $\rd \Pd/\rd t$ is calculated from a linear least-squares fit over a predetermined time window of length $\tfit$ before the current time, and the extrapolated disruptivity is restricted to the interval $\Pd \in \brack{0,1}$. \red{While} this is certainly not an optimized form of extrapolation, \sout{but} it \red{at least} provides a starting point from which to evaluate the survival analysis approach. Possible alternatives will be discussed in section~\ref{sec:future}.

Equation~\eqref{eq:S2} is powerful in that its evaluation provides an actual estimate for the probability of plasma survival (i.e. no disruption) a time $\Dt$ into the future, and it requires only a RF algorithm (well-)trained on binary-classified disruption data. Still, this approach does not actually trigger an alarm itself; we need to determine the threshold for mitigation or \soutt{other} \redd{specific} avoidance techniques. One possible alarm could be a threshold for the \emph{median time} $\mt$, defined by
    \begin{equation}
        S(t + \mt | t) = 0.5.
        \label{eq:median}
    \end{equation}
This is the time into the future beyond which \redd{the probability of} survival is less than 50\%. Another metric could be the \emph{expected \red{future} lifetime} $\tau$, \soutt{simply} calculated as
    \redd{
    \begin{eqnarray}
        \tau = \int_{0}^{\infty} S(t + t' | t) \, \rd t'.
        \label{eq:lifetime}
    \end{eqnarray}
    }\redd{It is worth noting that the expected \red{future} lifetime is usually (approximately) greater than or equal to the class time, i.e. $\tau \gtrsim \tc$, for cases when $\ts \ll \tc$; this is due to the ignorance introduced from binary classification \red{and a fast sampling rate which frequently updates the state of the plasma}. (See \ref{app:lifetime} for a derivation of this inequality \red{and further discussion}.)} For either case, if $\mt$ or $\tau$ \soutt{were less than} \redd{would reach} some threshold time, say $\tmin$ or $\tc$, then the disruption mitigation system could be triggered.\footnote{Note that for some specific cases, like $\Pd(t) = 0$, $\mt$ does not exist and $\tau$ diverges. In this paper, we assume that $\mt$ exists and $\tau$ is finite.} The median time and lifetime are used as examples throughout the rest of this paper, but choices of the most reliable metrics and optimization of thresholds are left for future work. 

Here, it is prudent to distinguish the present application of survival analysis from that in \cite{olofsson2018PPCF,olofsson2018FED}. In \cite{olofsson2018PPCF,olofsson2018FED}, a direct hazard model was used to calculate multivariable hazard functions from machine learning of experimental data, which could then be related to survival probabilities via \eqref{eq:int_h}. This is, in a way, opposite to the approach adopted in this paper; nevertheless, their model has benefits including the incorporation of time-dependent covariates. It is unclear, though, if these hazard functions can be calculated in real-time, which would be necessary for integration with a PCS.

\section{\redd{Application to test data}}\label{sec:test}

In this section, the Kaplan-Meier formalism for survival analysis is applied \redd{to test data} in two cases: First, \redd{\emph{smooth}} test data is used to demonstrate the capabilities and limitations of this approach, \redd{including calculations of the survival and hazard functions.} \redd{Second, an application to \emph{noisy} test data highlights} \soutt{highlighting} the effects of linear extrapolation, noise, and fitting windows. \soutt{Second,} \redd{Later, in section~\ref{sec:cmod},} the survival analysis approach is applied to \soutt{real} experimental data from the Alcator C-Mod tokamak to show its potential use in ``real time.''  

\subsection{\redd{Smooth data}}\label{sec:test_smooth}

\begin{figure}[h!]
    \centering
    \includegraphics[width=\myWidth]{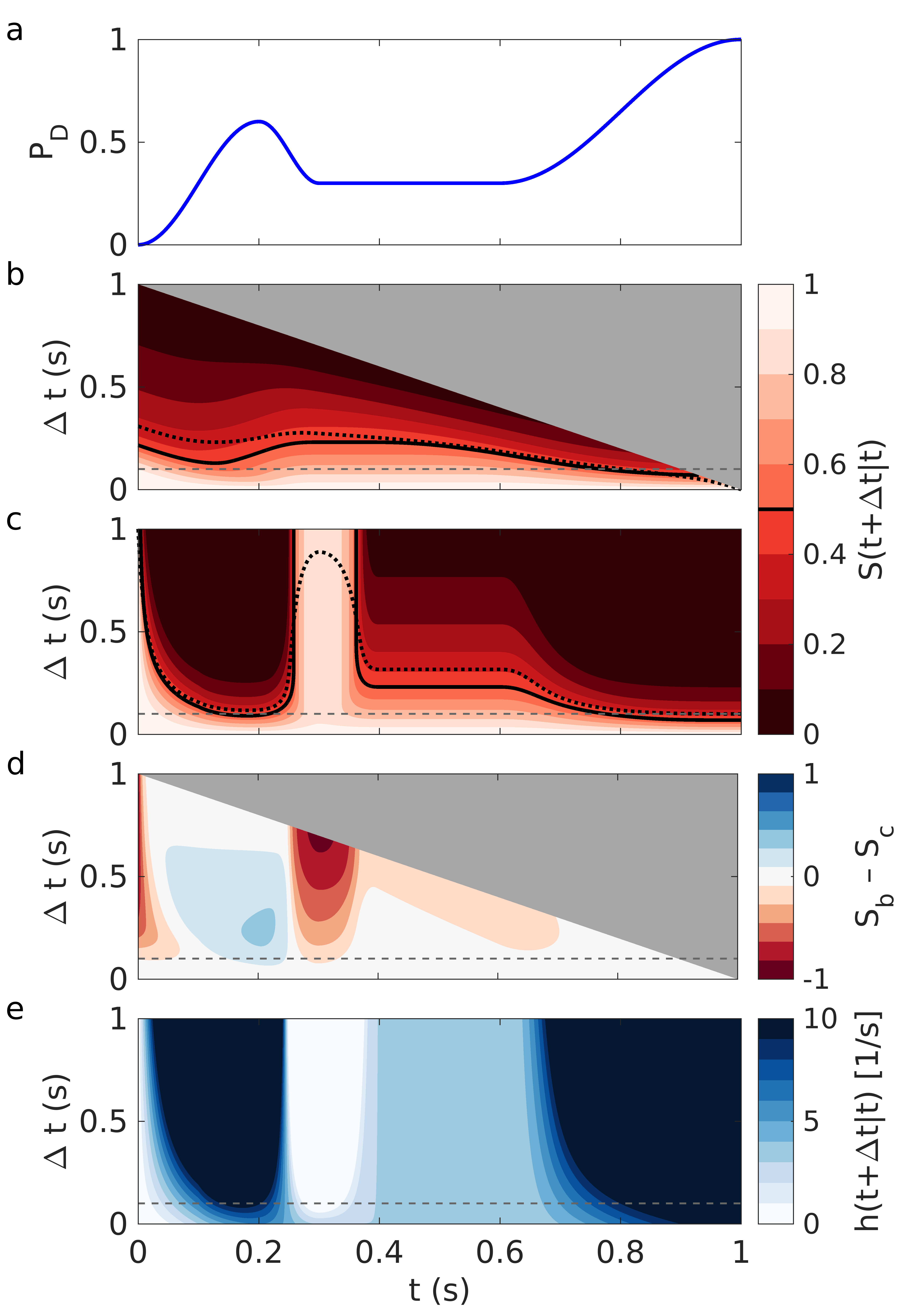}
    \caption{(a) A toy disruptivity signal for demonstration. Resulting contours of survival probability ($\tc$~=~100~ms, $\ts$~=~1~ms) for cases in which the future disruptivity is (b) known and (c) linearly-extrapolated ($\tfit$~=~100~ms). Median times (solid) and expected \red{future} lifetimes (dotted) overlay \soutt{subplots} (b) and (c). (d) Contours of the absolute difference in $S(t+\Dt|t)$ between \soutt{subplots} (b) and (c). (e) The hazard function computed from data in \soutt{subplot} (c). Grey regions in \soutt{subplots} (b) and (d) indicate unknown data. \redd{Horizontal dashed lines in (a)-(d) indicate $\Dt = \tc$.}}
    \label{fig:fakeSmooth}
\end{figure}

In this test case, a simple, ad hoc disruptivity signal was created and is shown in figure~\ref{fig:fakeSmooth}a. As is seen, there is an initial rapid increase in $\Pd$, peaking at $t$~=~0.2~s and then decaying to a steady-state value of $\Pd = 0.3$ from $t$~=~0.3-0.6~s. At $t$~=~0.6~s, the disruptivity signal increases gradually until reaching $\Pd = 1$ at $t$~=~1~s. \redd{Note that this does \emph{not} necessarily imply that a disruption would occur at $t$~=~1~s in this test case; instead, a disruption would be \sout{\emph{expected}} \red{\emph{likely}} to occur within the next class time, i.e. $t \in [1,1+\tc]$~s, if the disruptivity remained at $\Pd = 1$.}

Contours of survival probability, calculated from \eqref{eq:S2}, are shown in \redd{figures}~\ref{fig:fakeSmooth}b-c, with current time $t$ and time into the future $\Dt$ plotted on the horizontal and vertical axes, respectively. Here, a class time of $\tc$~=~100~ms and sampling time of $\ts$~=~1~ms were assumed. Each color corresponds to one decade within the range 0-100\%. In \soutt{subplot} \redd{figure}~\ref{fig:fakeSmooth}b, the future disruptivity signal is \emph{known} for all times; that is, this calculation predicts the future. Thus, data beyond $t + \Dt >$~1~s is unknown (grey region). Overlaid are contours of $\mt$ (solid) and \redd{the} \emph{effective} \red{future} lifetime (dotted), calculated as
    \begin{equation}
        \teff = \int_{0}^{1-t} S(t + t' | t) \, \rd t',
    \end{equation}
since data \sout{is} \red{are} only available up to $t$~=~1~s.

It is important to note several features of figure~\ref{fig:fakeSmooth}b that may seem counter-intuitive. First, even though $\Pd(t)$ is known for all $t$ (and $\Dt$), the contours vary in time. This is because the calculation ``domain'' changes at each $t$, and the probability of survival at the current time ($\Dt = 0$) is always 1. Also, one may have hoped that the effective \red{future} lifetime would always satisfy $t+\teff$~=~1~s since the calculation ``knows'' that $\Pd$($t$~=~1~s)~=~1. However, recall that a disruptivity value of 1 only conveys that a disruption \sout{\emph{should}} \red{is likely to} occur within the next time $\tc$. In fact, the value of survival probability at $t + \Dt$~=~1~s is not always zero. What is more, the data of figure~\ref{fig:fakeSmooth}b represents the best possible predictive capability of survival analysis applied to this particular test case. This is the ``ideal'' scenario against which we should compare more realistic scenarios.

Figure~\ref{fig:fakeSmooth}c shows contours of the survival function calculated using a linear extrapolation of $\Pd(t)$, as in \eqref{eq:taylor}, with fitting window $\tfit$~=~100~ms. Also overlaid are contours of $\mt$ and the \emph{approximate} \red{future} lifetime $\tapx$, calculated as
    \begin{equation}
        \tapx = \int_{0}^{1} S(t + t' | t) \, \rd t'.
        \label{eq:tapx}
    \end{equation}
Here, predictions are only made $\Dt$~=~1~s into the future, \redd{a limitation} which could be varied. This example highlights some drawbacks of the linear extrapolation method. The initial rise in disruptivity extrapolates to $\Pd$~=~1 around $t\approx$~0.3~s, causing both $\mt$ and $\tapx$ to dip \soutt{below} \redd{close to} $\tc$ \redd{around $t\approx$~0.2~s}. Then, the decrease in disruptivity from $t\approx$~0.2-0.3~s extrapolates to $\Pd$~=~0 around $t\approx$~0.4~s, at which time the plasma is predicted to survive far ($\Dt \geq$~1~s) into the future.

The absolute difference between survival functions in \soutt{subplots} \redd{figures}~\ref{fig:fakeSmooth}b and \ref{fig:fakeSmooth}c is given in \redd{figure}~\ref{fig:fakeSmooth}d, with magnitude in the range $[-1,1]$. While there is significant disagreement between the linearly-extrapolated and ``ideal'' cases around $t\approx$~0.3~s, most other times agree within $\pm$270~ms and some within $\pm$90~ms. Future work should explore how to best minimize these differences; some suggestions are made in section~\ref{sec:future}.

Finally, contours of the hazard function are plotted in figure~\ref{fig:fakeSmooth}e, calculated using the data in figure~\ref{fig:fakeSmooth}c and \eqref{eq:h}. As expected, the hazard is highest from $t$~=~0-0.2~s and 0.7-1.0~s and lowest from $t$~=~0.2-0.4~s. \redd{As derived in \ref{app:hazard}, the hazard function can take values approximately in the range $h \in [0,(\tc-\ts)^{-1}]$~s$^{-1}$; these failure rates correspond to predicted \red{future} lifetimes ranging from approximately $\tc$ to $\infty$ when $\ts \ll \tc$.} \soutt{Because the hazard function can take values in the range $[0, \infty]$~s$^{-1}$, it is difficult to intuit the physical meaning of $h(t+\Dt|t)$, and without any calibration} \redd{Because these ranges are dependent on the chosen class time, we must rely on comparing relative values. Future work should explore ``calibrating'' $h(t)$ and/or finding an appropriate combination of thresholds for both $S(t)$ and $h(t)$.}

\subsection{\redd{Noisy data}}\label{sec:test_noisy}

\begin{figure}[h!]
    \centering
    \includegraphics[width=\myWidth]{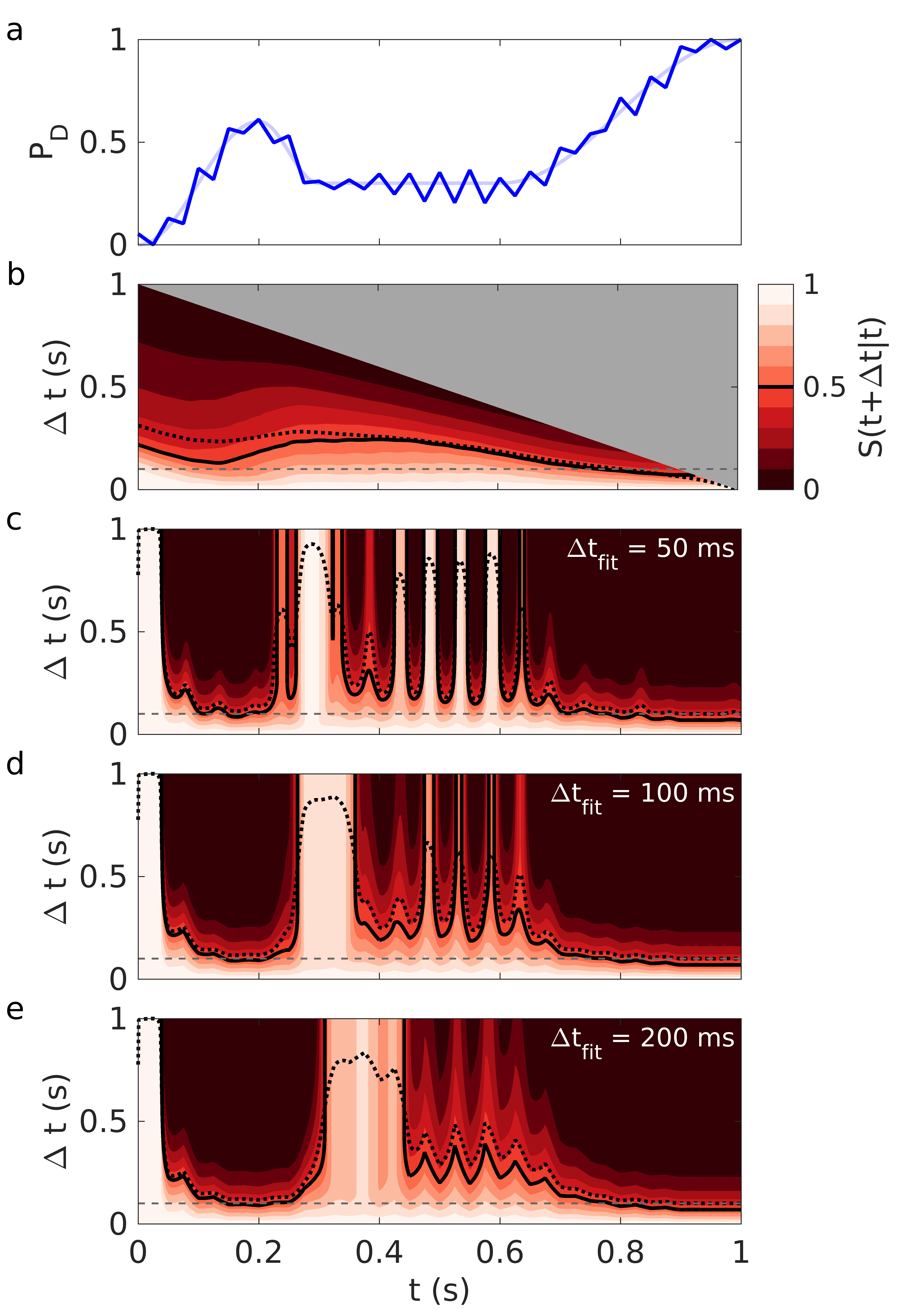}
    \caption{(a) The toy disruptivity data from figure~\ref{fig:fakeSmooth}a with added noise of period 50~ms. Resulting contours of survival probability ($\tc$~=~100~ms, $\ts$~=~1~ms) for cases in which the future disruptivity is (b) known and linearly-extrapolated with fitting time windows $\tfit$ of (c) 50~ms, (d) 100~ms, and (e) 200~ms. The median times (solid) and expected \red{future} lifetimes (dotted) are plotted. The color-scale in \soutt{subplot} (b) is the same for \soutt{subplots} (c)-(e). The grey region in \soutt{subplot} (b) indicates unknown data. \redd{Horizontal dashed lines in (b)-(e) indicate $\Dt = \tc$.}}
    \label{fig:fakeNoisy}
\end{figure}

The disruptivity signal used in the above analysis is smooth, unlike real data which will be noisy. Artificial noise, with random amplitude in the range $[-0.1,0.1]$ and with period (peak-to-peak) of 50~ms, was added to the smooth signal and is plotted in figure~\ref{fig:fakeNoisy}a. \redd{These variations in $\Pd$ are similar to those observed in the disruptivity calculated for real C-Mod data, as seen in figures~\ref{fig:good}-\ref{fig:false}.} The survival probability, calculated assuming the future disruptivity is \emph{known}, is shown in figure~\ref{fig:fakeNoisy}b; for low amplitude noise, it is (unsurprisingly) quite similar to that in figure~\ref{fig:fakeSmooth}b.

\soutt{Subplots} \redd{Figures}~\ref{fig:fakeNoisy}c-e show contours of the survival function computed using linear extrapolation, from \eqref{eq:taylor}, with fitting windows of $\tfit$~=~50, 100, and 200~ms, respectively. The median times and approximate \red{future} lifetimes, from \eqref{eq:tapx}, are also overlaid. Note, as seen in \soutt{subplot} \redd{figure}~\ref{fig:fakeNoisy}c, how \redd{using} a time window on the order of the period of noise oscillations \soutt{is} greatly \redd{affects $S(t)$} \soutt{affected by the noise}, with $\mt$ and $\tapx$ varying widely on this timescale. As the time window increases, the contours smooth out in time $t$, and variations in $\mt$ and $\tapx$ decrease. However, the ``responsiveness'' to changes on the order of $\tfit$ is also delayed; this is observed as both a rightward shift and broadening of the light-colored feature ($\Pd \geq 0.7$) in \soutt{subplots} \redd{figures}~\ref{fig:fakeNoisy}c-e as $\tfit$ increases from 50 to 200~ms. Future work should consider optimization of the fitting time window as well as proper uncertainty quantification. Here, it appears that \redd{a value of} $\tfit$ between 100-200~ms, i.e. 2-4 times the noise period, would best approximate the smooth data in figure~\ref{fig:fakeSmooth}c.

\section{\redd{Application to Alcator C-Mod data}}\label{sec:cmod}

\soutt{The Alcator C-Mod tokamak is a high magnetic field ($\Bo$~=~2-8~T), compact ($\Ro$~=~68~cm, $a$~=~22~cm) device with plasma currents and densities of order 1~MA and $10^{20}$~m$^{-3}$, respectively. A database of disruption parameters has been compiled for over four thousand C-Mod discharges, containing over $10^5$ data points in total \cite{rea2018ppcf,montes2019}.} \redd{As mentioned in section~\ref{sec:intro}, the} Random Forest algorithm\soutt{s} \redd{DPRF has} \soutt{have} been trained and tested on C-Mod data \cite{rea2018ppcf,montes2019}, with most recent results optimizing an \emph{alarm window} to trigger mitigation, as reported in \cite{montes2019}: If the disruptivity increases above a \emph{high threshold} without decreasing below a \emph{low threshold} over a \emph{time window}\footnote{Note that this \emph{alarm} time window \redd{$\tW$} is different from the \emph{fitting} time window $\tfit$ discussed elsewhere in the present paper.} \redd{$\tW$}, an alarm is triggered. \redd{In this work, the optimized DPRF algorithm had high and low threshold values of $\Pd$~=~0.35 and 0.05, respectively, an alarm window $\tW$~=~5~ms, and a class time $\tc$~=~325~ms. See \cite{montes2019} for more details on the DPRF algorithm optimization.}

\redd{This section investigates three C-Mod discharges on which this threshold-based alarm system has been \emph{tested} in ``real time.'' The first, in section~\ref{sec:good}, is a \emph{good prediction}, or true positive, meaning that an alarm would have been triggered with sufficient time to mitigate (or avoid) the impending disruption. The second, in section~\ref{sec:late}, is an example of a \emph{late warning}, or false negative, for which an alarm is triggered with too little time to mitigate. Finally, section~\ref{sec:false} presents a case of a \emph{false positive}; i.e. an alarm would have been triggered for a non-disrupting discharge. An example of a \emph{true negative}---when a non-disrupting discharge is correctly identified---is not included in the present work because disruptivity values $\Pd \approx 0$ do not illuminate any additional capabilities of the survival analysis approach.}

\subsection{\redd{A good prediction}}\label{sec:good}

\begin{figure}[h!]
    \centering
    \includegraphics[width=\myWidth]{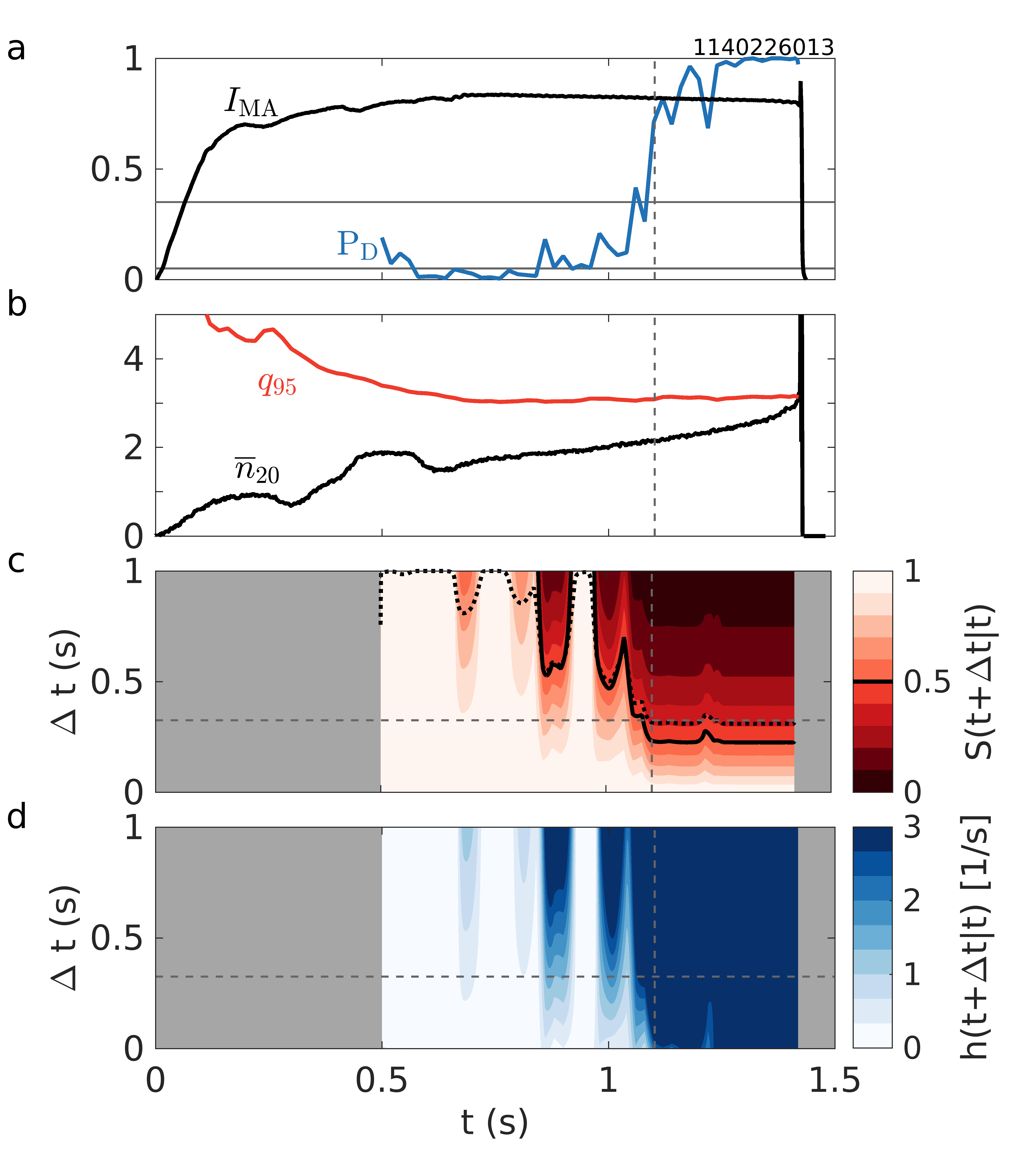}
    \caption{\redd{A good prediction: The (a) plasma current in MA and disruptivity and (b) edge safety factor and line-averaged plasma density in $10^{20}$~m$^{-3}$ are plotted for Alcator C-Mod discharge \#1140226013. Contours are shown for the (c) survival and (d) hazard functions, calculated using linear extrapolation as in \eqref{eq:taylor} ($\tfit$~=~100~ms, $\ts$~=~1~ms). The estimated \red{future} lifetime $\tapx$ (dotted) and median time $\mt$ (solid) are given in (c). The horizontal solid lines in (a) indicate the low and high thresholds of the RF algorithm, as described in the text. The dashed lines in (a)-(d) indicate $\tc$~=~325~ms before the disruption (vertical) and $\Dt = \tc$ (horizontal). Grey regions indicate unknown data.}}
    \label{fig:good}
\end{figure}

Figures~\ref{fig:good}a-b show plasma parameters for C-Mod discharge \#1140226013, during which the density increased steadily and toroidal magnetic field (not shown) decreased in time from $\Bo$~=~5.3~T to 3.9~T over $t\approx$~0-0.7~s. Since the plasma current was held constant at $\Ip$~=~0.8~MA, the edge safety factor \soutt{thus} decreased to $q_{95}\approx$~3. \redd{A locked mode, identified by a reduction in sawteeth observed in the plasma temperature (not shown), began around $t\approx$~1~s and \sout{is} \red{was} likely the cause of the disruption at $t\approx$~1.4~s.} \soutt{At first glance,} The disruptivity signal in figure~\ref{fig:good}a seems to accurately predict the disruption: $\Pd$ remains low during the first part of the flattop current ($t\approx$~0.5-1.0~s) with $\Pd$ increasing after $t\approx$~1~s and reaching $\sim$1 approximately 200~ms before the disruption. \soutt{However, For this optimized RF predictor,} \redd{The optimized DPRF predictor, discussed above, performs well on this discharge. The calculated disruptivity} $\Pd$ crosses the high threshold \redd{$\Pd = 0.35$} and remains above the low threshold \redd{$\Pd = 0.05$} \redd{(shown as solid horizontal lines in figure~\ref{fig:good}a)} over the \redd{required} time window \redd{$\tW$~=~5~ms} before the class time $\tc$~=~325~ms, indicated by the vertical dashed line \soutt{in figure~\ref{fig:good}}. \soutt{See \cite{montes2019} for more details on this RF algorithm optimization.}

The application of survival analysis to this experimental data gives another perspective of the disruptivity's predictive capability. Figures~\ref{fig:good}c-d show the survival and hazard functions, \soutt{from \eqref{eq:S2} and \eqref{eq:h} respectively,} using linear extrapolation from \eqref{eq:taylor} and a fitting time window $\tfit$~=~100~ms. Note that the disruptivity was actually calculated with two sampling times distinguished by proximity to the disruption time~$\td$: $\ts$~=~20~ms when $\td - t >$~20~ms, and $\ts$~=~1~ms when $\td-t \leq$~20~ms.\footnote[6]{These sampling rates were chosen to reduce the total database size and were based on considerations unrelated to the present study. The database can be updated to vary/increase the sampling rate.} Here, it is assumed that a real-time sampling rate of $\ts$~=~1~ms is achievable, so $\Pd(t)$ was interpolated appropriately. The median time and approximate \red{future} lifetime overlay the survival probability in \soutt{subplot} \redd{figure}~\ref{fig:good}c. \soutt{Note that both $\mt$ and $\tapx$ decrease} \redd{Note that $\mt$ falls} below $\Dt = \tc$, plotted also as a horizontal dashed line in \soutt{subplot} figure~\ref{fig:good}c, \redd{and $\tapx$ approaches $\tc$} around the time that $\td - t = \tc$. In fact, $\tapx \approx \tc$ during almost all times $\td - t < \tc$. This is \soutt{primarily} due to the linear extrapolation of $\Pd$ to 1 around this time. Thus, we conclude that the survival function has ``accurately'' predicted the disruption in this discharge. In this scenario, we could imagine setting an alarm with a requirement that $\mt$ \soutt{or $\tapx$} remains below $\tc$ \redd{(or $\tapx \approx \tc$)} for a certain time interval, perhaps a few confinement times ($\tE \sim$~20-30~ms in C-Mod). The hazard function is shown for completeness in figure~\ref{fig:good}d. As expected, $h(t)$ increases in amplitude as $\mt$ and $\tapx$ decrease\redd{, reaching a maximum at $h = (\tc - \ts)^{-1}\approx$~3.1~s$^{-1}$}. \soutt{However, the maximum value is $\sim$3 for this C-Mod data, whereas the maximum value was $\sim$10 for the presented test case. This further motivates a calibration of the hazard function---discussed in section~\ref{sec:summary}---to find typical value ranges for different machines and plasma parameters.}

\subsection{\redd{A late warning}}\label{sec:late}

\begin{figure}[h!]
    \centering
    \includegraphics[width=\myWidth]{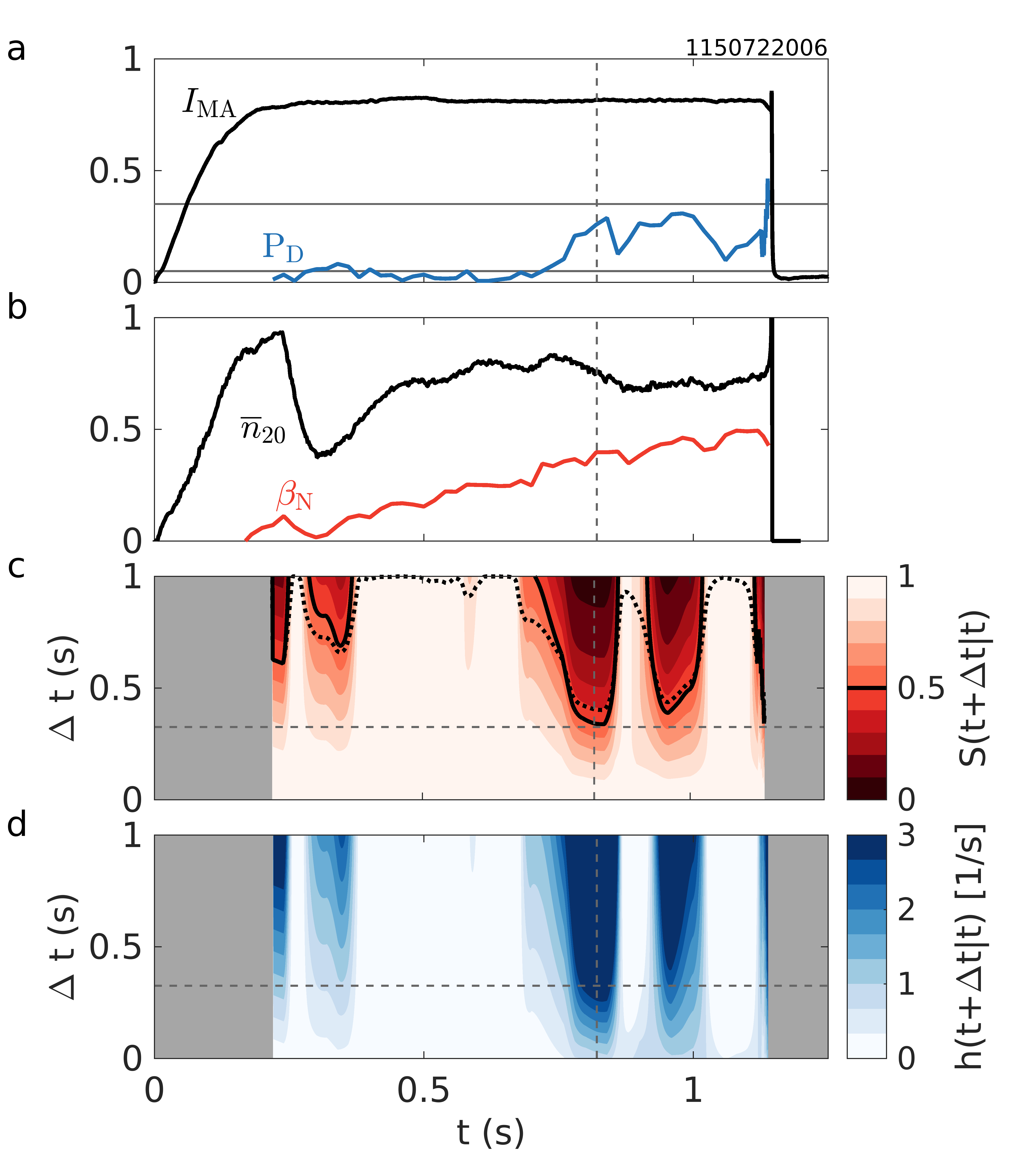}
    \caption{\redd{A late warning: The (a) plasma current in MA and disruptivity and (b) normalized plasma $\beta$ (\%) and line-averaged plasma density in $10^{20}$~m$^{-3}$ are plotted for Alcator C-Mod discharge \#1150722006. Contours are shown for the (c) survival and (d) hazard functions, calculated using linear extrapolation as in \eqref{eq:taylor} ($\tfit$~=~100~ms, $\ts$~=~1~ms). The estimated \red{future} lifetime $\tapx$ (dotted) and median time $\mt$ (solid) are given in (c). The horizontal solid lines in (a) indicate the low and high thresholds of the RF algorithm, as described in the text. The dashed lines in (a)-(d) indicate $\tc$~=~325~ms before the disruption (vertical) and $\Dt = \tc$ (horizontal). Grey regions indicate unknown data.}}
    \label{fig:late}
\end{figure}

\redd{
Plasma parameters for C-Mod discharge \#1150722006, which ended in a disruption at $t \approx$~1.15~s, are shown in figures~\ref{fig:late}a-b. The disruptivity signal $\Pd$ output from the DPRF algorithm remained low during much of the flattop current, i.e. $t \approx$~0.2-0.7~s. During this time, the plasma $\beta$ was steadily increasing, while the edge safety factor (not shown) decreased. As the plasma density decreased \red{from} $t \approx$~0.7-0.9~s, $\Pd$ increased slightly, although not enough to pass the high threshold $\Pd = 0.35$ and trigger an alarm. Finally, only $\sim$20~ms before the disruption, $\Pd$ rose rapidly above the threshold; this likely resulted from an impurity injection, which was observed as a fast increase in radiated power (not shown). Unfortunately, the disruptivity passed the threshold $\Pd > 0.35$ too close to the disruption, qualifying this as a false negative; that is, the DPRF failed to predict the disruption with enough time to mitigate or avoid it.
}


\redd{
For this discharge, the survival and hazard functions, calculated using a fitting window $\tfit$~=~100~ms and sampling time $\ts$~=~1~ms, are plotted in figure~\ref{fig:late}c-d. As is seen, neither the median time $\mt$ nor approximate \red{future} lifetime $\tapx$ fall below the class time $\Dt = \tc$ at any point during the discharge. Thus, the predictive capabilities of survival analysis would fail to adequately warn of an impending disruption in this case. In fact, unlike a disruptivity-threshold scheme, the survival analysis framework may have not triggered an alarm at all since both $\mt$ and $\tapx$ are greater than $\tc$, even near the disruption time. More work must be done to optimize this framework. 
}


\subsection{\redd{A false alarm}}\label{sec:false}

\begin{figure}[h!]
    \centering
    \includegraphics[width=\myWidth]{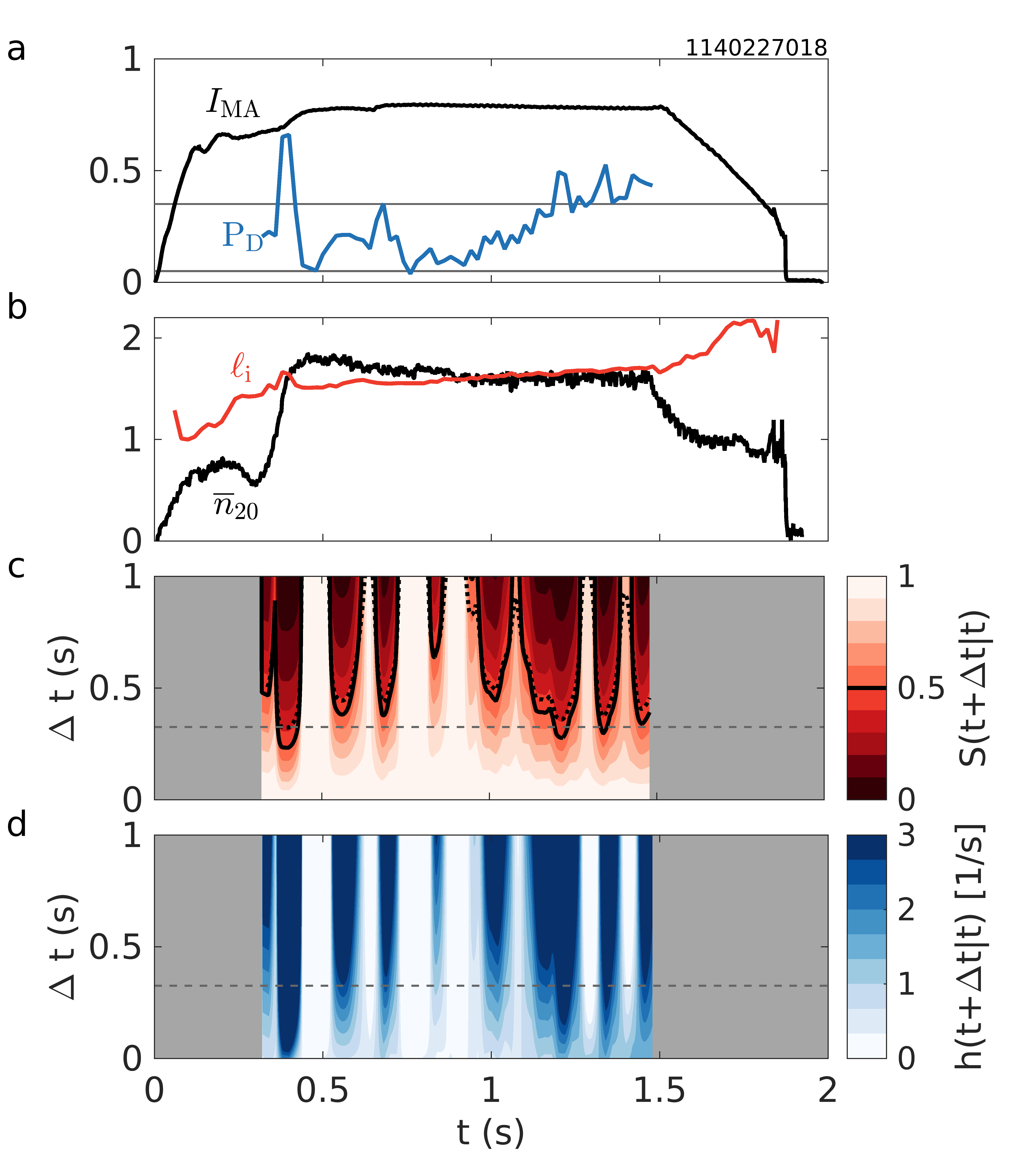}
    \caption{\redd{A false alarm: The (a) plasma current in MA and disruptivity and (b) normalized internal inductance and line-averaged plasma density in $10^{20}$~m$^{-3}$ are plotted for Alcator C-Mod discharge \#1140227018. Contours are shown for the (c) survival and (d) hazard functions, calculated using linear extrapolation as in \eqref{eq:taylor} ($\tfit$~=~100~ms, $\ts$~=~1~ms). The estimated \red{future} lifetime $\tapx$ (dotted) and median time $\mt$ (solid) are given in (c). The horizontal solid lines in (a) indicate the low and high thresholds of the RF algorithm, as described in the text. The horizontal dashed lines in (c) and (d) indicate $\Dt = \tc$. Grey regions indicate unknown data.}}
    \label{fig:false}
\end{figure}

\redd{
Figures~\ref{fig:false}a-b show plasma parameters for C-Mod discharge \#1140227018. During this discharge, tokamak operators applied error fields to purposefully achieve a locked mode. However, the plasma did not lock and was successfully ramped down without disrupting, as seen in the time traces of plasma current and density. The DPRF algorithm generated a disruptivity signal which has a large spike during the current ramp-up and later a steady increase from $\Pd \approx$~0-0.5 over $t \approx$~0.75-1.5~s. Considering only the flattop portion of the discharge, the rise in $\Pd$ was likely due to the peaking of the current profile, seen as an increase in internal inductance, as well as magnetic pickup from the error field coils. Because $\Pd$ crossed the high disruptivity threshold of $\Pd$~=~0.35 at $t \approx$~1.2~s, this discharge qualifies as a false positive; that is, in the proposed alarm-threshold scheme, this discharge would be identified as disruptive even though it does not actually disrupt. 
}

\redd{
However, using the estimated \red{future} lifetime threshold suggested above, the survival analysis framework correctly classifies this discharge. The survival and hazard functions, computed using the linear extrapolation method with fitting window $\tfit$~=~100~ms and sampling time $\ts$~=~1~ms, are shown in figures~\ref{fig:false}c-d. The data are quite noisy, with large variations in median time $\mt$ and approximate \red{future} lifetime $\tapx$. A larger fitting time window, as discussed in section~\ref{sec:test_noisy}, could improve this. Note that $\mt$ falls below the class time $\tc$ at $t \approx$~1.2 and 1.35~s, the same times as the peaks in $\Pd$ seen in figure~\ref{fig:false}a. However, the estimated \red{future} lifetime $\tapx$ never decreases to $\tc$ during the flattop portion of the discharge. Of course, a wide variety of thresholds for $\mt$ and $\tapx$ could be chosen such that a false alarm \emph{would} \red{have been} \sout{be} triggered by this survival analysis approach. Nevertheless, physics intuition might tell us that if the expected \red{future} lifetime exceeds the class time chosen for disruptive data, then an alarm should not be sounded.
} 

\section{\redd{Summary}}\label{sec:summary}

In this paper, we proposed a survival analysis approach to disruption prediction, applying the Kaplan-Meier formalism to existing Random Forest (RF) algorithms trained on binary-classified data. While it is unknown whether this method could predict disruptions with higher accuracy or more warning time than traditional machine learning (ML) methods, we argued that survival analysis provides a more physically-meaningful and more easily-interpretable framework for disruption prediction. 

This work required a database of binary-classified data, split into non-disruptive and disruptive classes by a class time $\tc$ before the disruption. Given that a plasma state was disruptive, we assumed that the time of the impending disruption was uniformly distributed. In addition, the output ``disruptivity'' signal $\Pd(t)$ from a Random Forest (RF) algorithm was interpreted as the probability that the plasma state at time $t$ belongs to the disruptive class. Therefore, at time $t_i$, knowledge of $\Pd(t_i)$ and $\tc$ allows evaluation of the probability of a disruption in a future time interval $\Dt$.

The Kaplan-Meier formalism, from survival analysis, was then used to calculate the survival probability $S(t)$ from a \emph{time-evolving} disruptivity signal. At any time $t$, the survival function $S(t+\Dt|t)$ actually gives the probability of survival (i.e. no disruption) a time $\Dt$ into the future. From this, the median time $\mt$ and (approximate) future lifetime $\tapx$ were used as indicators of plasma ``health,'' where low values of either $\mt$ or $\tapx$ could be used to trigger an alarm of an impending disruption.

Application of this model to test data illuminated some non-intuitive, probabilistic features of the survival function---including $\mt, \tapx > 0$ for $\Pd(t) = 1$---and drawbacks of linearly-extrapolating $\Pd(t)$ to predict future disruptivity values. Moreover, the hazard function $h(t)$, which indicates the instantaneous rate-of-failure, was found to be consistent with expectations: low values of $h(t)$ for high values of $\mt$ and $\tapx$, and vice versa. However, $h(t)$ \soutt{must be calibrated, or} \redd{is dependent on the choice of $\tc$, so} comparison of relative values may only be useful. When noise was added to the smooth test disruptivity signal, $S(t)$ proved to be quite sensitive to the length of \redd{the} ``fitting'' time window $\tfit$ used for linear extrapolation. If such a method is used in future analyses, this time window should be several ($\sim$2-4) times longer than the longest period of noise oscillations.

This survival analysis methodology was also applied to real disruption data from \soutt{one} \redd{three} Alcator C-Mod discharges. The disruptivity signal was generated from \soutt{an RF} \redd{the DPRF} algorithm trained on the C-Mod disruption database and optimized with $\tc$~=~325~ms. \soutt{As expected, both $\mt$ and $\tapx$} \redd{In one discharge, the median time $\mt$} fell below $\tc$ \redd{and the expected \red{future} lifetime $\tapx$ approached $\tc$} for times within $\tc$ of the disruption time $\td$; that is, \redd{$\mt,\tapx \lesssim \tc$} for $\td - t \leq \tc$\redd{, as hoped.} For this specific scenario, the application of survival analysis to disruption prediction should be considered successful. \redd{However, in another case, both the disruptivity threshold scheme and survival analysis approach were not able to adequately warn, with sufficient time, of an impending radiative collapse caused by an impurity injection. For the final discharge, the optimized DPRF disruptivity signal $\Pd$ and thresholds falsely triggered an alarm for a non-disrupting plasma; the approximate \red{future} lifetime $\tapx$ from survival analysis, however, remained above the class time $\tc$, indicating that such a false alarm could have been avoided.}

\section{\redd{Opportunities for future work}}\label{sec:future}

\redd{The survival analysis approach presented in this paper has several limitations, and there is much work to be done before this framework could be implemented in real time in a plasma control system.} Future work \soutt{should identify these cases,} \redd{must} explore a variety of metrics \redd{(i.e. beyond $\mt$ and $\tapx$)} and optimize thresholds and durations required to trigger mitigation or pursue avoidance strategies. For instance, perhaps a predictor would be more successful by monitoring the third-quartile time, \redd{i.e.} $S(t+t_{75}|t) = 0.75$, and triggering an alarm if $t_{75}$ decreases below the minimum time \redd{$\tmin$} required for mitigation \soutt{$\tmin$}.

\redd{
Here, the authors would like to comment on some specific limitations of this analysis and suggest several opportunities for improvement:
}
\begin{enumerate}
    \item\label{item:one} In this paper, we assumed the existence of an RF disruption prediction algorithm already trained on disruption-relevant covariates. However, the field of survival analysis has many robust statistical methods for calculating survival probabilities, among other quantities, which incorporate time-dependent covariates. See, for instance, the Cox proportional hazards model \cite{cox1972} or a more general ``direct'' hazard model \cite{olofsson2018PPCF,olofsson2018FED}. These methods should be implemented for disruption prediction and compared to current ML algorithm performance. \redd{In addition,} other established fields, like probabilistic risk assessment, should \sout{also} be explored as alternative methods for disruption prediction and avoidance.
    \item The analysis in this work relied on the assumption of a well-calibrated disruptivity signal; that is, \redd{a value of} $\Pd = 0$ or 1 \soutt{truly meant} \redd{was assumed to truly indicate} that the plasma state resides in the non-disruptive or disruptive class, respectively. Future work with ML algorithms should take care to properly calibrate output signals, for instance as described in \cite{olofsson2018PPCF,olofsson2018FED,niculescu2005}. \redd{This approach might also be applied to other ML methods trained on two classes and with output signals in the range $[0,1]$, such as those described in \cite{yoshino2003,cannas2004,windsor2005,murari2008,aledda2015,murari2018}.}
    \item Additionally, the physical meaning of ML algorithm output signals must be well understood. Fortunately for RF algorithms trained on two-class data, this interpretation is relatively straightforward. However, the ``distance to a boundary'' as calculated by a Support Vector Machine model, for example, might not be as clearly interpreted.
    \item \redd{Another assumption of this analysis was a uniform probability density of times of data in the disruptive class. However, in principle, the probability density could be learned from the database itself. One extension of this work would be to consider more carefully the sampling rate of data in the database and/or to adjust the ML algorithm to remove any related biases.}
    \item Finally, the linear extrapolation of disruptivity employed in this paper is quite \redd{simple} \redd{(even crude)}, especially considering that plasma parameters can be actuated in real time to navigate in ``disruptivity space.'' A smarter method, such as that described in \cite{parsons2017}, could be used to calculate gradients in parameter space and map \sout{out} possible trajectories away from the disruptive boundary ($\Pd = 1$, in this case). \red{Furthermore, dynamical models of the the plasma state vector $\mathbf{x}(t)$---which is input into the disruption predictor, i.e. $\Pd[\mathbf{x}(t)]$---could be implemented to more realistically extrapolate the future disruptivity.}
\end{enumerate}

\section*{Acknowledgements}

The authors thank E. Olofsson and M. Parsons for enlightening discussions, as well as the entire Alcator C-Mod team. \redd{Comments from the reviewers also strengthened this paper.} This work was supported by US DOE Grant DE-FC02-99ER54512, using Alcator C-Mod, a DOE Office of Science User Facility.

\redd{R. Sweeney is supported by the U.S. Department of Energy Fusion Energy Sciences Postdoctoral Research Program administered by the Oak Ridge Institute for Science and Education (ORISE) for the DOE. ORISE is managed by Oak Ridge Associated Universities (ORAU) under DOE contract number DE-SC0014664. All opinions expressed in this paper are the authors' and do not necessarily reflect the policies and views of DOE, ORAU, or ORISE.}

\appendix

\redd{
\section{Calculation of the minimum expected \red{future} lifetime}\label{app:lifetime}
In this section, it is shown that the expected \red{future} lifetime $\tau$, as calculated from \eqref{eq:lifetime}, does not fall far below the class time $\tc$, chosen to bifurcate data into non-disruptive and disruptive classes, for most cases of interest. Recall that the probability of survival beyond time $t+t'$, assuming survival until time $t$, is given by \eqref{eq:S2}
    \begin{equation}
        S(t + t'|t) = \prod_{j=0}^{n} \brack{1 - \Pd\paren{t + \frac{jt'}{n}} \frac{t'}{n\tc} },
        \label{eq:S_appA}
    \end{equation}
where the future time $t'$ has been split into $n$ steps. As described in section~\ref{sec:survival}, equation~\eqref{eq:S_appA} is the Kaplan-Meier estimator of survival probability, where the probability of failure (i.e. a disruption) is intuited to be the product of the disruptivity signal $\Pd(t)$ at each time step and the fractional time interval $t'/n\tc$ from \eqref{eq:P1}. It is important to emphasize again that the time step must be less than or equal to the class time, i.e. $t'/n \leq \tc$, for this relation to hold. This can be achieved by the appropriate choice of $n$ for given $t'$ and $\tc$; in practice, the sampling time $\ts = t'/n$ of a disruption predictor is often much shorter than $\tc$ anyway. 
}

\redd{
Because the survival function is always non-negative ($S(t) \geq 0$) and monotonically-decreasing ($\rd S/\rd t \leq 0$), the expected \red{future} lifetime, from \eqref{eq:lifetime}, satisfies 
    \begin{equation}
        \tau = \int_{0}^{\infty} S(t + t' | t)  \rd t' \geq \int_{0}^{T} S(t + t' | t) \rd t',
        \label{eq:lifetime_appA}
    \end{equation}
where $T$ is assumed finite. Before using \eqref{eq:S_appA} as the integrand of \eqref{eq:lifetime_appA}, note that $S(t)$ is minimal for $\Pd(t) = 1$; therefore, we can write
    \begin{equation}
        \tau \geq \int_{0}^{T} \prod_{j=0}^{n} \brack{1 - \frac{t'}{n\tc}} \rd t' = \int_{0}^{T} \paren{1 - \frac{t'}{n\tc}}^{n+1} \rd t'.
        \label{eq:lifetime2_appA}
    \end{equation}
Here, it is important to note that (i) the integral is over the domain $t' \in [0, T]$ and (ii) $t'/n \leq \tc$ must still be satisfied. Thus, the upper limit is maximally $T = n\tc$. \emph{If} we wanted to evaluate the integral for $T > \tc$, we would need to use a different form of \eqref{eq:S_appA} in accordance with \eqref{eq:P1}, but that is not necessary here. The integral of \eqref{eq:lifetime2_appA} is evaluated to be
    \begin{equation}
         \tau \geq -\frac{n}{n+2} \tc \left. \paren{1 - \frac{t'}{n\tc}}^{n+2}\right\vert_0^{n\tc} = \frac{n}{n+2} \tc.
    \end{equation}
For the lowest possible sampling rate, $n = 1$ and then $\tau \geq \tc/3$ \red{for $t' \leq \tc$}. However, for most cases of interest, $n\gg 1$ since $\ts \ll \tc$; then we arrive at the desired result $\tau \gtrsim \tc$.}

\red{As a final remark, note that the survival function of \eqref{eq:S_appA} approaches the exponential $\exp(-t'/\tc)$ in the limit $n \to \infty$ when $\Pd(t) = 1$. In this limit, the first integral of \eqref{eq:lifetime_appA} can be carried out explicitly so that the expected future lifetime is exactly $\tau = \tc$. It may seem counter-intuitive that the calculated future lifetime is \emph{not} less than the class time, i.e. $\tau < \tc$, when the plasma is always predicted to be in the disruptive state, i.e. when $\Pd = 1$. However, this is a consequence of (i) the uncertainty associated with a choice of class time $\tc$ in binary classification of the data sets, (ii) a sampling time much shorter than $\tc$ which provides frequent updates of the current plasma state, and (iii) the framework adopted in this paper which treats probabilities between time steps as independent. (Refer to sections~\ref{sec:disruption} and \ref{sec:survival} for further discussion.) Future work can explore the inclusion of past data in the prediction of future survival. For instance, perhaps a prediction of $\Pd = 1$ over the past $\tc$ seconds should automatically trigger an alarm. In the end, $\tau$ may not even be the most appropriate metric, or quantity to monitor, in successive studies since the survival probability at future time $\tau$ can often be less 50\%, i.e. $S(t+\tau|t) < 0.5$, as seen in figures~\ref{fig:fakeSmooth}-\ref{fig:false}.}


\redd{
\section{Calculation of the maximum hazard}\label{app:hazard}
In this section, the (approximate) maximum value of the hazard function is calculated. From \eqref{eq:h}, the hazard function is computed
    \begin{equation}
        h(t) = -\frac{\rd \ln S(t)}{\rd t}.
    \end{equation}
Once again, the survival function is given by
    \begin{equation}
        S(t) = \prod_{j=0}^{n} \brack{1 - \Pd\paren{\frac{jt}{n}} \frac{t}{n\tc} },
        \label{eq:S_appb}
    \end{equation}
where it is assumed that the initial time is $t = 0$ and $t/n \leq \tc$. Taking the logarithm of \eqref{eq:S_appb} turns the product into a summation
    \begin{equation}
        \ln S(t) = \sum_{j=0}^{n} \ln \brack{1 - \Pd\paren{\frac{jt}{n}} \frac{t}{n\tc} }.
        \label{eq:lnS_appB}
    \end{equation}
Here, we utilize the linear extrapolation from \eqref{eq:taylor}
    \begin{equation}
        \Pd\paren{\frac{jt}{n}} \approx \Pdo + \frac{jt}{n}\dPodt,
        \label{eq:Pd_appB}
    \end{equation}
where $\Pdo = \Pd(0)$ and $\rd\Pdo/\rd t = \rd\Pd(0)/\rd t$ are evaluated at the starting time, and the range of values is restricted to $\Pd \in [0, 1]$. Substituting \eqref{eq:Pd_appB} into \eqref{eq:lnS_appB} and taking the derivative gives
    \begin{equation}
        h(t) = \sum_{j=0}^{n} \brack{\frac{\Pdo}{n\tc} + \frac{2jt}{n^2\tc}\dPodt} \brack{ 1 - \Pd\paren{\frac{jt}{n}} \frac{t}{n\tc}}^{-1}.
    \end{equation}
This relation is maximal for $\Pd(jt/n) \to 1$. Taking this limit and evaluating the summation gives
    \begin{equation}
        h(t) \leq \frac{1}{\tc} \frac{n+1}{n} \paren{\Pdo + t\dPodt } \paren{1 - \frac{t}{n\tc}}^{-1}.
        \label{eq:h_appb}
    \end{equation}
Because $\Pd \to 1$, it follows that the middle term of \eqref{eq:h_appb} is (maximally) 1, so that 
    \begin{equation}
        h(t) \leq \frac{n+1}{n} \paren{\tc - \frac{t}{n}}^{-1}.
    \end{equation}
For the lowest sampling rate, $n = 1$ and then $h(t) \leq 2/(\tc - t)$, which can approach infinity for $t \to \tc$. However, for more realistic cases with a sampling time $\ts = t/n$ and $n \gg 1$, we find
    \begin{equation}
        h(t) \lesssim \frac{1}{\tc - \ts}.
    \end{equation}
}

\section*{References}

\end{document}